\documentclass[journal, 11pt, onecolumn, draftclsnofoot]{IEEEtran}
\usepackage[usenames, dvipsnames]{color}
\usepackage[left=1in, right=1in, top=1.25in, bottom=1.25in]{geometry}
\usepackage{citesort}
\usepackage{setspace}
\usepackage{pifont}
\newcommand{\cmark}{\ding{51}}%
\ifCLASSINFOpdf
  \usepackage[pdftex]{graphicx}
	\usepackage{epstopdf}
\else
  \usepackage[dvips]{graphicx}
\fi
\usepackage{moresize}
\usepackage[cmex10]{amsmath}
\usepackage{algorithmic}
\usepackage{amssymb}
\usepackage{array}
\usepackage{mdwmath}
\usepackage{mdwtab}
\usepackage{eqparbox}
\usepackage{multirow}
\usepackage{fixltx2e}
\usepackage{enumerate}
\usepackage{url}
\usepackage{tabularx}
\usepackage{verbatim}
\usepackage{color}
\usepackage{soul}

\newcommand{\add}[1]{{\color{black}#1}}
\newcommand{\chg}[1]{{\color{black}#1}}

\newcommand{\Cmb}{\mathbb{C}}
\newcommand{\Rmb}{\mathbb{R}}

\newcommand{\mbf}[1]{\boldsymbol{#1}}

\newcommand{\sumi}{\sum_{i=1}^{N}}

\newcommand{\sumk}{\sum_{k=1}^{K}}

\newcommand{\Rf}{\mathfrak{R}}

\newcommand{\xb}{\mbf{x}}

\newcommand{\yb}{\boldsymbol{y}}

\newcommand{\Db}{\mbf{D}}

\newcommand{\Wb}{\mbf{W}}

\renewcommand{\IEEEbibitemsep}{0pt plus 0.5pt}
\makeatletter
\IEEEtriggercmd{\reset@font\normalfont\fontsize{7.7pt}{8.20pt}\selectfont}
\makeatother
\IEEEtriggeratref{1}

\begin{document}
\title{Transform Learning for Magnetic Resonance Image Reconstruction: From Model-based Learning to Building Neural Networks}
\author{Bihan~Wen,~\IEEEmembership{Member,~IEEE},~Saiprasad~Ravishankar,~\IEEEmembership{Member,~IEEE},\\Luke~Pfister,~\IEEEmembership{Student Member, IEEE},~and~Yoram~Bresler,~\IEEEmembership{Life Fellow,~IEEE}\vspace{-0.4in}
\thanks{DOI: 10.1109/MSP.2019.2951469. Copyright (c) 2019 IEEE. Personal use of this material is permitted. However, permission to use this material for any other purposes must be obtained from the IEEE by sending a request to
pubs-permissions@ieee.org}
\thanks{\scriptsize B. Wen is with the School of Electrical and Electronic Engineering at Nanyang Technological University, Singapore, 639798 email: bihan.wen@ntu.edu.sg.}
\thanks{\scriptsize S. Ravishankar is with the Departments of Computational Mathematics, Science and Engineering, and Biomedical Engineering at Michigan State University, East Lansing, MI, 48824 USA email: ravisha3@msu.edu.}
\thanks{\scriptsize L. Pfister and Y. Bresler are with the Department of Electrical and Computer Engineering and the Coordinated Science Laboratory, University of Illinois, Urbana-Champaign, IL, 61801 USA e-mail: (lpfiste2, ybresler)@illinois.edu. Their work was supported in part by the National Science Foundation
under Grant IIS 14-47879. The work Of L. Pfister was also supported in part by 
the National Cancer Institute of the National
Institutes of Health under Award Number R33CA196458. The content is solely the responsibility
of the authors and does not necessarily represent the official views of the National Institutes of
Health.}}
\maketitle

\vspace{-0.2in}
\begin{abstract}
Magnetic resonance imaging (MRI) is widely used in clinical practice, but it has been traditionally limited by its slow data acquisition. Recent advances in compressed sensing (CS) techniques for MRI reduce acquisition time while maintaining high image quality. Whereas classical CS assumes the images are sparse in known analytical dictionaries or transform domains, methods using learned image models for reconstruction have become popular. The model could be pre-learned from datasets, or learned simultaneously with the reconstruction, i.e., blind CS (BCS). Besides the well-known synthesis dictionary model, recent advances in transform learning (TL) provide an efficient alternative framework for sparse modeling in MRI. TL-based methods enjoy numerous advantages including exact sparse coding, transform update, and clustering solutions, cheap computation, and convergence guarantees, and provide high-quality results in MRI compared to popular competing methods. This paper provides a review of some recent works in MRI reconstruction from limited data, with focus on the recent TL-based methods. A unified framework for incorporating various TL-based models is presented. We discuss the connections between transform learning and convolutional or filterbank models and corresponding multi-layer extensions, with connections to deep learning. Finally, we discuss recent trends in MRI, open problems, and future directions for the field.
\end{abstract}

\begin{IEEEkeywords}
\vspace{-0.1in}
Sparse signal models, Convolutional models, Transform learning, Dictionary learning, Structured models, Compressed sensing, Machine learning, Physics-driven deep learning, Multi-layer models, Efficient algorithms, Nonconvex optimization, Magnetic resonance imaging, Computational imaging.
\end{IEEEkeywords}

\vspace{-0.15in}
\section{Introduction} \label{section1}

Magnetic resonance imaging (MRI) is a widely used imaging modality in routine clinical practice. \add{It is noninvasive, nonionizing, and offers a variety of contrast mechanisms and excellent visualization of both anatomical structure and physiological function.} However, a traditional limitation of MRI affecting both throughput (scan time) and image resolution, especially in dynamic imaging, \add{is that it is a relatively slow imaging technique because the measurements
are acquired sequentially over time.}

\begin{figure}[!t] 
\begin{center} 
\vspace{-0.15in}
\begin{tabular}{ccc}
\includegraphics[width=1.2in]{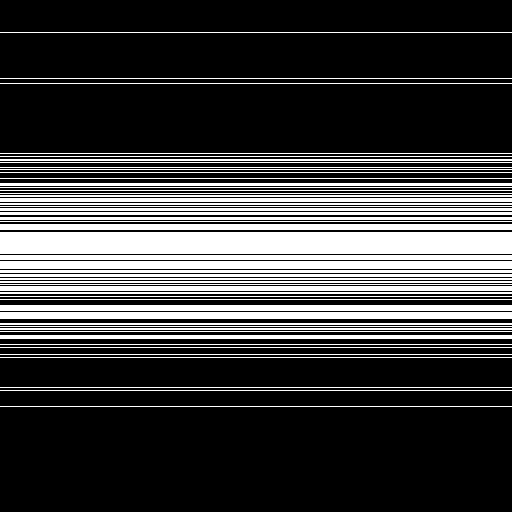} &
\includegraphics[width=1.17in]{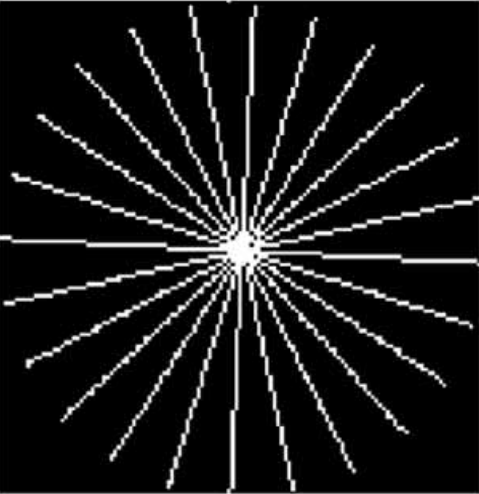} &
\includegraphics[width=1.2in]{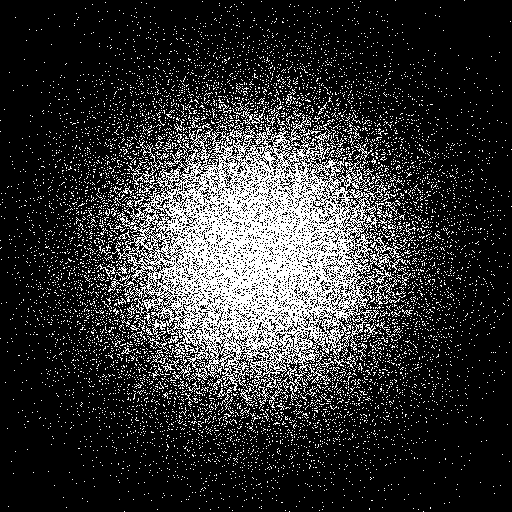} \\
 Cartesian & Radial & 2D random \\
\end{tabular} 
\caption{\add{Examples of under-sampling in k-space using Cartesian, Radial (from~\cite{lingal13}), and 2D random patterns. Schemes such as 2D random or pseudo-radial~\cite{sai2011dlmri} sampling are feasible when data corresponding to multiple image slices are jointly acquired and the frequency encode direction is perpendicular to the image plane.}} \label{fig:mask} 
\end{center} 
\vspace{-0.5in} 
\end{figure}

Recent advances in MRI include improved pulse sequences for rapid acquisition, ultra-high field imaging for improved signal-to-noise ratio, and hardware-based parallel data acquisition (P-MRI) methods~\cite{pMRI-Survey,setsompop16}. P-MRI enables acquiring fewer Fourier, or k-space, samples by exploiting the diversity of multiple RF receiver coils, and is widely used in commercial systems and clinical applications.
\add{Compressed Sensing (CS) methods}
\add{\cite{Fen-PT97,donoho2006compressed}} have also been successfully applied to MRI~\cite{lustig2007sparse} to significantly reduce the number of samples and corresponding acquisition time needed for accurate image reconstruction. CS theory enables the recovery of images from significantly fewer measurements than the number of unknowns by assuming  sparsity of the image in a known transform domain or dictionary, and requiring the acquisition to be appropriately incoherent with the transform; albeit at the cost of a nonlinear
reconstruction procedure.
In practice, CS-based MRI methods typically use variable density random sampling schemes during acquisition \cite{lustig2007sparse} (see Fig.~\ref{fig:mask}) along with sparsifying models such \add{as wavelets and finite difference operators.
In 2017, the FDA approved the use of CS-based MRI in clinical practice.}


While early CS MRI methods exploited sparsity in analytical dictionaries and transform domains, recent years have seen growing interest in \textit{learning} the underlying MR image models for reconstruction \cite{sai2011dlmri}. The models may be learned from a corpus of data, or jointly with the reconstruction (i.e., blind compressed sensing)~\cite{sai2011dlmri,sravTCI1}. The latter approach provides high data-adaptivity, but requires more complex and typically highly nonconvex optimization. \add{Recent methods even train iterative learning-based algorithms in a supervised manner using training pairs of ground truth and undersampled data~\cite{sun2016deep,ravcfess17,hammernik2018learning}.}
In this review paper, we first discuss \add{early sparsity and low-rank model-based techniques for CS MRI}, followed by 
\add{later}
advances, 
\add{particularly}
in learning-based methods for MRI reconstruction.
We focus 
\add{mainly}
on \textit{sparsifying transform learning} (TL) based reconstruction \cite{sabres13,sravTCI1} models and schemes, \add{which offer numerous advantages such as cheap computations; exact sparse coding, clustering and other updates in algorithms; convergence guarantees; ease in incorporating a variety of model properties and invariances; and effectiveness in reconstruction.} Importantly, \add{these methods also produce state-of-the-art results in applications~\cite{wen2018power,wensaibresvidosat19}}, under a common umbrella. 
We review various TL-based methods and models under a unified framework, and illustrate their promise over \add{some} competing methods.
We also consider the connections of TL methods and multi-layer extensions, to neural networks and discuss recent trends, open questions, and future directions in the field.
The goal of this paper is not to provide a comprehensive review of all classes of MRI reconstruction methods, but rather to focus on the recent transform 
learning
\add{class of techniques and elucidate their properties, underlying models, benefits, connections, and extensions.} 

\add{The rest of this article is organized as follows. Section~\ref{section2} reviews sparsity and low-rank based CS MRI approaches, followed by learning-based methods including TL-based schemes.} Section~\ref{section3} provides a tutorial on TL-based MRI. Section~\ref{section4} discusses interpretations of transform learning-based methods and extensions, along with new research directions and open problems. We conclude in Section~\ref{section5}.

\vspace{-0.15in}
\section{CS MRI Reconstruction: From Nonadaptive Methods to Machine Learning}
\label{section2}

\begin{figure}[!t] 
\begin{center} 
\vspace{-0.25in}
\begin{tabular}{c}
\hspace{-0.18in}\includegraphics[width=6in]{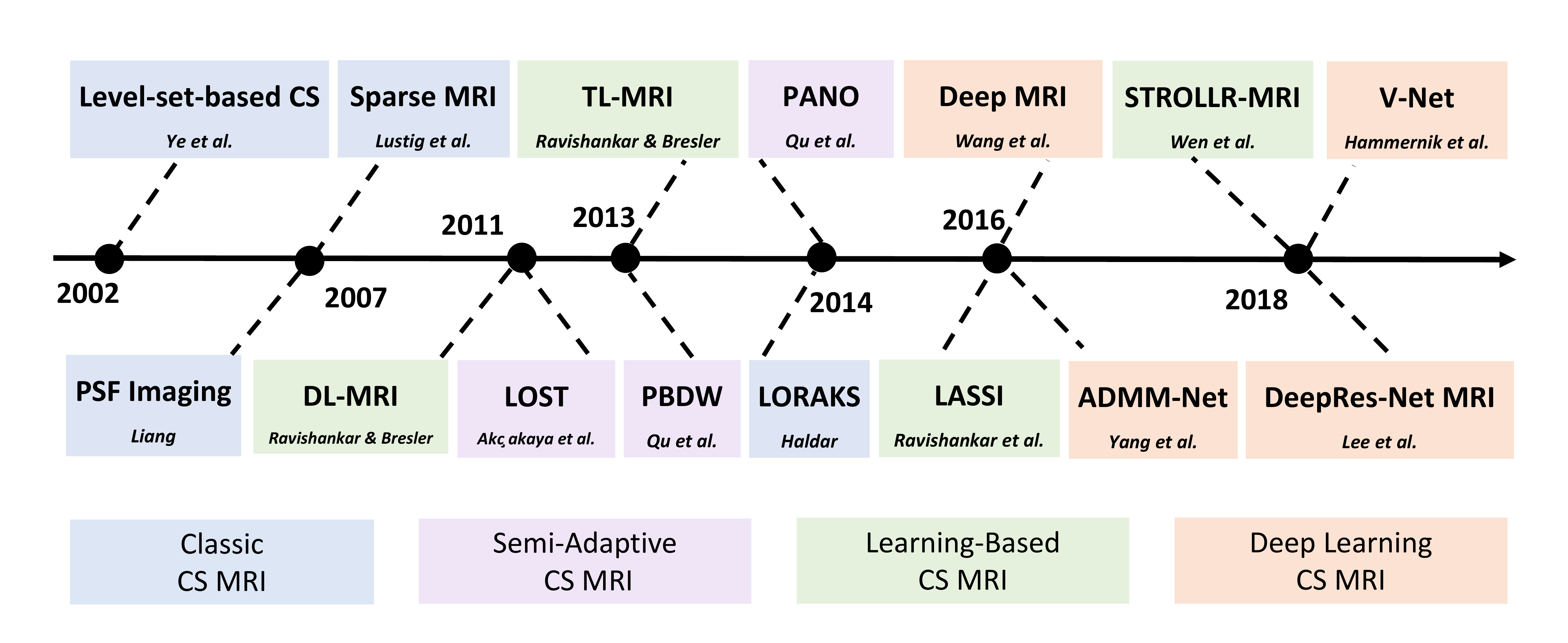} \\
\end{tabular} 
\vspace{-0.16in} 
\caption{\add{Timeline for evolution of Classical CS MRI (with analytical models) to recent learning-based CS MRI methods. Only limited papers are included as examples among each class of methods (categories are not strictly chronological).}} \label{fig:timeline} 
\end{center} 
\vspace{-0.4in} 
\end{figure}

MRI reconstruction from limited measurements is an ill-posed inverse problem, and thus effective models or priors on the underlying image are necessary for accurate image reconstruction.
CS MRI methods use random sampling techniques that create incoherent or noise-like aliasing artifacts when the conventional (zero-filling) inverse FFT reconstruction is used. Image models and \add{corresponding penalty functions (i.e., 
regularizers),} such as those based on sparsity, are used to  effectively remove the artifacts during \add{reconstruction.} 
This section surveys some of the progress in MRI reconstruction from \add{limited or CS} data starting with \add{early 
approaches based on analytical sparsity and low-rankness,}
followed by recent advances in learning-based MRI reconstruction. 
\add{A timeline for evolution of classical CS MRI to learning-based CS MRI in past years, with some representative works in each class is presented in Fig.~\ref{fig:timeline}.}

\vspace{-0.15in}
\subsection{Sparsity and Low-rank Models in MRI}

Early CS MRI approaches assumed that MR images are sparse
under analytical transforms~\cite{ye2002self,donoho2006compressed}, such as wavelets~\cite{lustig2007sparse}, 
contourlets, or total variation (TV)~\cite{lustig2007sparse}.
Later works incorporated more sophisticated models into the reconstruction framework.
Examples include exploiting self-similarity of MR images via Wiener filtering to improve reconstruction quality~\cite{akccakaya2011low},
the balanced sparse model for tight frames~\cite{liu2015balanced,liu2016projected}, and the Patch-Based Directional Wavelets (PBDW)~\cite{qu2012undersampled} and PAtch-based Nonlocal Operator (PANO)~\cite{qu2014magnetic} methods that use semi-adaptive wavelets and are thus more flexible than traditional wavelets.
\add{Low-rank data models have also been used for MRI reconstruction
such as the Partially Separable Functions (PSF)~\cite{liang2007spatiotemporal} approach.} 
In dynamic MRI, where the measurements are inherently undersampled, 
\add{low-rank models that exploit the temporal correlation of the dynamic image sequence are popular.  
More recent (also see~\cite{ravmoorerajfes17}) low-rank based methods include the annihilating filter-based low-rank Hankel matrix (ALOHA) approach and the LORAKS scheme~\cite{haldar2014low}.}









\vspace{-0.1in}
\subsection{Data-driven or Learning-Based Models for Reconstruction}
\vspace{-0.05in}

Learning-based methods for MRI have shown promising improvements over nonadaptive schemes. 
\add{The early dictionary-blind CS MRI method, dubbed DL-MRI~\cite{sai2011dlmri} used dictionary learning (DL) as a data-driven regularizer} for MRI reconstruction to achieve significantly improved results over previous nonadaptive schemes.
DL-MRI learned a \add{small} patch-based 
\add{synthesis} dictionary while simultaneously performing image reconstruction, thus the model is highly adaptive to the underlying \add{object or patient.}



\add{However, each iteration of DL-MRI involved
\emph{synthesis sparse coding} using a greedy algorithm, which is computationally expensive.
Unlike the synthesis dictionary model 
that approximates image patches as sparse linear combinations of columns of a dictionary, i.e., an NP-hard sparse coding problem, the complementary sparsifying transform model assumes that the image patches are approximately sparse in a transform (e.g., wavelet) domain.}
A key advantage of this framework is that, unlike synthesis sparse coding, \emph{transform domain sparse coding} is a simple thresholding operation \cite{sabres13}.
Recent transform learning (TL) based reconstruction schemes include efficient, closed-form updates in the iterative algorithms~\cite{ravishankar2015efficient,sravTCI1}.
TL models are closely tied to convolutional models \cite{pfisbres19,saibrenmulti}. 
Several TL-MRI schemes have shown promise for MRI including STL-MRI~\cite{ravishankar2015efficient} that learns a square and non-singular transform operator, UNITE-MRI~\cite{sravTCI1} that learns a union of transforms with 
a clustering step,
FRIST-MRI~\cite{wen2017frist} that learns a large union of transforms related by rotations,
and STROLLR-MRI~\cite{wen2018power} that combines low-rank modeling and block matching with transform learning. The latter models can be viewed as hybrid models.

\add{Recent works have also developed efficient synthesis dictionary learning-based reconstruction schemes such as SOUP-DIL MRI \cite{ravrajfes17}, and LASSI \cite{ravmoorerajfes17} that uses a low-rank + learned dictionary-sparse model for dynamic MRI.
The most recent trend involves supervised (e.g., deep) learning of MRI models such as those based on convolutional neural networks \cite{wang2016accelerating,schlemper18,leeye18,hammernik2018learning,wang2018image}.
Some of these works incorporate the measurement forward model (physics) in the reconstruction model that is typically an unrolled iterative algorithm \cite{ravcfess17,sun2016deep,hammernik2018learning}. Supervised learning of TL-MRI models has also shown promise
\cite{ravcfess17,cfess18}.}


In this paper, we focus 
on TL-MRI methods, which offer flexibility and enjoy numerous modeling, computational, convergence, and performance benefits.


\vspace{-0.2in}
\subsection{Qualitative Comparison of Different Methods}

Table~\ref{Tab:compareMRI} presents a qualitative comparison of a sample set of 
methods
in terms of the models and techniques they exploit.
While the 
Sparse MRI method \cite{lustig2007sparse} used a fixed sparsifying model, later works exploited directional features (e.g., \chg{PBDW~\cite{qu2012undersampled}, or its recent extension FDLCP~\cite{zhan2016fast} that grouped patches with their directionality and learned corresponding orthogonal dictionaries}), block matching (e.g., PANO \cite{qu2014magnetic}), and low-rank modeling (e.g., LORAKS). Methods such as DLMRI, STL-MRI, FRIST-MRI, STROLLR-MRI, SOUPDIL-MRI, ADMM-Net \cite{sun2016deep}, BCD-Net~\cite{ravcfess17,ravacfess18,cfess18}, LASSI~\cite{ravmoorerajfes17}, etc., \add{all involve model learning.}

\setlength{\tabcolsep}{.25em}
\renewcommand{\arraystretch}{0.80}
\begin{table*}[t]
\vspace{-0.2in}
\centering
\fontsize{10}{16pt}\selectfont
\begin{tabular}{|c|c|c|c|c|c|c|c|}
\hline
\multirow{2}{*}{\textbf{ Methods }} & \multicolumn{4}{c|}{Sparse Model} & Block & Supervised & Low-Rank  \\
\cline{2-5}
 & Fixed & Directional & DL & TL  & Matching & Learning & Modeling\\
\hline
Sparse MRI~\cite{lustig2007sparse} & \cmark & & & & & & \\
\hline
PBDW~\cite{qu2012undersampled} & \cmark & \cmark &  & & & & \\
\hline
LORAKS~\cite{haldar2014low} &  &  & & &  & & \cmark \\
\hline
PANO~\cite{qu2014magnetic} & \cmark  &  & &  & \cmark & & \\
\hline
DLMRI~\cite{sai2011dlmri} & & & \cmark & & & & \\
\hline
SOUPDIL-MRI~\cite{ravrajfes17} & & & \cmark & & & & \\
\hline
LASSI~\cite{ravmoorerajfes17} & & & \cmark & & & & \cmark \\
\hline
STL-MRI~\cite{ravishankar2015efficient} & & &  & \cmark & & & \\
\hline
FRIST-MRI~\cite{wen2017frist} & &  \cmark & & \cmark & & & \\
\hline 
STROLLR-MRI~\cite{wen2018power} & & &  & \cmark & \cmark & & \cmark  \\
\hline
ADMM-Net~\cite{sun2016deep} & & &  &  &  & \cmark & \\
\hline
BCD-Net~\cite{ravcfess17,cfess18} & & &  & \cmark  &  & \cmark & \\
\hline
\end{tabular}
\caption{\chg{Comparison between several types of MR image reconstruction methods surveyed in this work.} }
\label{Tab:compareMRI}
\vspace{-0.6in}
\end{table*}

\vspace{-0.15in}
\section{Tutorial on Transform Learning-based MRI}
\label{section3}
Transform learning schemes have been shown to be effective for MR image reconstruction from limited measurements \cite{ravishankar2015efficient,sravTCI1,wen2017frist,wen2018power}. 
As a variety of TL-MRI algorithms have been proposed, each based on different transform models and learning schemes, it is important to understand:
\begin{enumerate}
\item What are the relationships and differences among the TL-MRI schemes? 
\item What MR image properties are used in each transform model? 
\item Which methods are most effective for reconstruction of a particular MR image?
\end{enumerate}
To this end, we present a tutorial that is intended to unify all recent TL-MRI schemes, and summarizes their problem formulations and algorithms using a general framework. 
\add{We discuss and contrast the features of several TL-MRI schemes, namely STL-MRI~\cite{ravishankar2015efficient}, UNITE-MRI~\cite{sravTCI1}, FRIST-MRI~\cite{wen2017frist}, and STROLLR-MRI~\cite{wen2018power}, and visualize their learned models.
We also illustrate the benefits of TL-MRI using STROLLR-MRI as compared to other classes of MRI reconstruction methods.}



\vspace{-0.2in}
\subsection{\add{CS-MRI Formulation}}



Given the k-space measurements $\yb \in \Cmb^m$ of the (vectorized) MR image $\xb \in \Cmb^p$, \add{the theory of Compressed Sensing~\cite{donoho2006compressed}} 
enables accurate image recovery provided that $\xb$ is sufficiently sparse in some transform domain, and the sampling of $\yb$ is incoherent with the sparsifying transform.
In an ideal case without measurement noise, \add{a simple formulation of the CS reconstruction problem is the following:}
\vspace{-0.1in}
\begin{equation} 
\nonumber (\mathrm{P0})\;\;\;\;\;\;\;\;\; \hat{\mbf{x}} = \underset{\mbf{x}}{\operatorname{argmin}} \: \left \| \, \Psi \, \xb \,  \right \|_0 \, \;\text{s.t.} \;\; \mbf{F}_u \mbf{x} = \mbf{y}. 
\label{eq:CSMRI}
\vspace{-0.1in}
\end{equation}
Here, $\mbf{F}_u \in \Cmb^{m \times p}$ (with $m \ll p$) denotes the under-sampled Fourier encoding matrix~\cite{ravishankar2015efficient}, which is the sensing or measurement operator in MRI. For P-MRI, the measurement operator also incorporates sensitivity (SENSE) maps \cite{pMRI-Survey}.
When sampling on a Cartesian grid with a single coil, $\mbf{F}_u \triangleq \mbf{U} \mbf{F}$ where $\mbf{U} \in \Rmb^{m \times p}$ is a down-sampling matrix (of zeros and ones), and $\mbf{F} \in \Cmb^{p \times p}$ is the full
Fourier encoding matrix normalized such that $\mbf{F}^H \mbf{F} = \mbf{I}_p\,$, where $\mbf{I}_p \in \Rmb^{p \times p}$ is the identity matrix.
Fig.~\ref{fig:mask} displays three 
\add{k-space undersampling masks.}
Matrix $\Psi \in \Cmb^{p \times p}$ is a sparsifying transform  and $\xb$ is assumed sparse in the $\Psi$-transform domain. 
The $\ell_0$ 
``norm" is a sparsity measure that counts the number of nonzeros in a vector.
Alternative sparsity promoting functions include $\ell_p$ ($0 < p < 1$) 
penalties or the convex $\ell_1$ norm penalty. 
The goal in $(\mathrm{P0})$ is to seek the $\Psi$-domain sparsest solution $\hat{\xb}$ that satisfies the imaging \add{forward} model $\mbf{F}_u \mbf{x} = \mbf{y}$. 
\add{In MRI, since $\yb$ is usually noisy, 
the reconstruction problem is typically formulated with a data-fidelity penalty as follows:}
\vspace{-0.15in}
\begin{equation} 
\nonumber (\mathrm{P1})\;\;\;\;\;\;\;\;\; \hat{\mbf{x}} = \underset{\mbf{x}}{\operatorname{argmin}} \:  \left \| \, \Psi \, \xb \,  \right \|_0 + \upsilon  \left \| \mbf{F}_u \mbf{x} - \mbf{y} \right \|_{2}^{2}\, .
\label{eqn:CS-MRI2}
\end{equation}
\add{Here, the $\ell_2$ data fidelity term $ \left \| \mbf{F}_u \mbf{x} - \mbf{y} \right \|_{2}^{2}$ with $\upsilon>0$ is based on Gaussian measurement noise.}

\vspace{-0.1in}
\subsection{\add{General TL-MRI Framework and Its Variations}}

\add{In practice, there are multiple limitations in using $(\mathrm{P0})$ or $(\mathrm{P1})$ for MR image reconstruction:}
\begin{itemize}
    \item \add{Instead of exact sparsity in the transform domain, MR images are typically only approximately sparse.}
    \item The transform $\Psi$ 
    is pre-defined and fixed. It is not optimized for the underlying image(s) $\xb$.
    \item 
    Instead of imposing common sparsity properties for the entire image, it may be more effective to assume local or nonlocal diversity or variability of the models.
\end{itemize}

\begin{figure}[!t] 
\begin{center} 
\vspace{-0.15in} 
\begin{tabular}{c}
\includegraphics[width=6.4in]{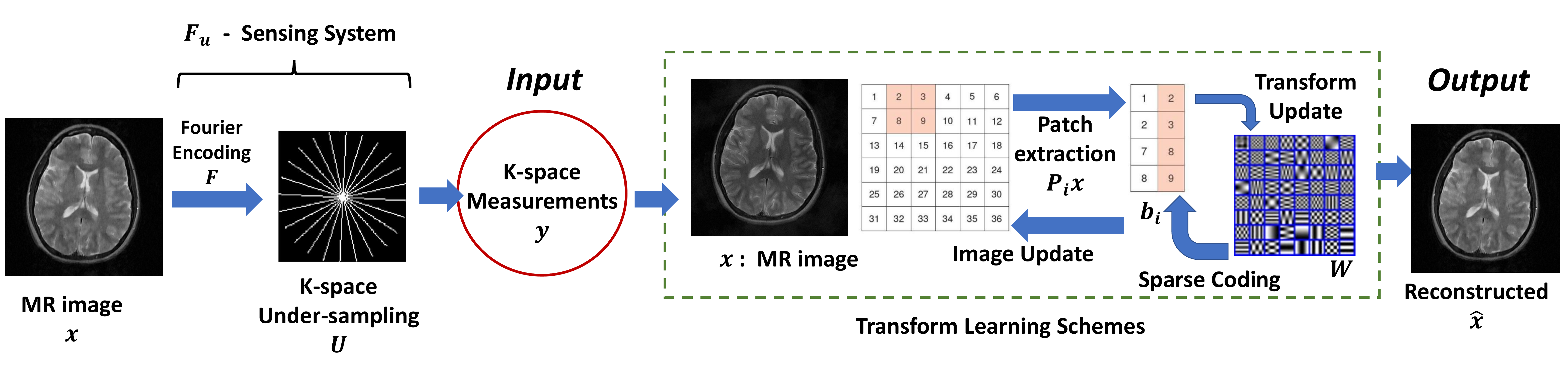} 
\end{tabular} 
\vspace{-0.2in} 
\caption{\add{A general pipeline of MR image reconstruction with sparsifying transform learning.}} \label{fig:tl-patch} 
\end{center} 
\vspace{-0.35in} 
\end{figure}


\add{Recent TL-MRI works~\cite{ravishankar2015efficient,sravTCI1,wen2017frist,wen2018power} addressed these limitations by adapting an \emph{approximately} sparsifying transform model to the MR image, and incorporating rich sparsifying models of image content. 
These formulations  can be written in the following unified form:} 
\begin{equation} 
\nonumber (\mathrm{P2})\;\;\;\;\;\;\;\;\; \hat{\mbf{x}} = \underset{\mbf{x}}{\operatorname{argmin}} \:  \left \| \mbf{F}_u \mbf{x} - \mbf{y} \right \|_{2}^{2} + \Rf_{TL} (\mbf{x})\, ,
\label{framework:TL-MRI}
\end{equation}
where the functional \add{$\Rf_{TL} (\mbf{x})$ is a transform-learning based regularizer.} 
\add{The actual form of $\Rf_{TL} (\mbf{x})$ depends on the underlying image properties and models, and is the major difference between the various TL-MRI formulations.
Another difference in the form of the regularizer arises from whether the transforms were learned from a training set or learned directly during reconstruction. The latter approach, 
involving optimization over both the image and the model parameters, 
is called \emph{blind compressed sensing} (BCS). One could also learn the transform from a training set and use it and the image reconstructed with it to initialize BCS algorithms to adapt the model to the specific data.}
In short, one can plug in the desired TL into $(\mathrm{P2})$ through $\Rf_{TL} (\mbf{x})$, and Problem $(\mathrm{P2})$ will reduce to the corresponding variation of the general TL-MRI scheme. Fig.~\ref{fig:tl-patch} is a general pipeline for \add{a BCS} TL-MRI scheme.

We review several 
recent TL-MRI schemes 
such as STL-MRI~\cite{ravishankar2015efficient}, UNITE-MRI~\cite{sravTCI1}, FRIST-MRI~\cite{wen2017frist}, and STROLLR-MRI~\cite{wen2018power}.
We discuss how they can be incorporated in the general framework $(\mathrm{P2})$ by using specific \add{transform models and learned regularizers, and also discuss the properties they exploit. We discuss the methods under the more common BCS setup.}
Several of the TL-MRI algorithms above also have proven convergence guarantees to critical points of the underlying problems~\cite{ravishankar2015efficient,sravTCI1,wen2017frist}. 

\subsubsection{STL-MRI~\cite{ravishankar2015efficient}}
The earliest formulation of TL-MRI applied square transform learning (STL)~\cite{ravishankar2015efficient} for MR image reconstruction, dubbed STL-MRI.
The regularizer $\Rf_{TL} (\mbf{x}) = \Rf_{STL} (\mbf{x})$ is defined as
\begin{align}  
\Rf_{STL} (\mbf{x}) \triangleq 
\; \underset{\mbf{W}, \{ \mbf{b}_i \}}{\operatorname{argmin}} \sumi \{ \left \| \Wb \mathbf{P}_i \xb - \mbf{b}_i \right \|_2^2 + \tau^2 \left \| \mbf{b}_i \right \|_0 \}  + \frac{\lambda}{2} \left \| \Wb \right \|_F^2 - \lambda \log ( \, \det \Wb  \, )\; .
\label{eq:STLMRI}
\end{align}
Here and in the remainder of this work, when certain indexed variables are enclosed within braces, it represents the set of all variables over the range of the indices, e.g., $\{ \mbf{b}_i \}$ in $(\ref{eq:STLMRI})$ represents $\{ \mbf{b}_i \}_{i=1}^N$.
The operator $\mathbf{P}_i \in \Rmb^{n \times p}$ extracts a $\sqrt{n} \times \sqrt{n}$ square patch (block) from the image in vectorized form as $\mathbf{P}_i \xb \in \Cmb^n$ \add{(see Fig.~\ref{fig:tl-patch}).} 
We assume $N$ patches in total, and the square transform $\Wb \in \Cmb^{n \times n}$ is 
assumed to sparsify $ \mathbf{P}_i \xb $, with the transform sparse approximation denoted as $\mbf{b}_i$.
The last two terms in \eqref{eq:STLMRI} are the regularizers for the transform that enforce useful properties on the transform during learning. 
\add{Here,
 the $- \log ( \, \det \Wb  \, )$ penalty prevents trivial solutions (e.g., $\Wb = \mathbf{0}$ or with repeated rows), and the $\left \| \Wb \right \|_F^2$ penalty prevents a scale ambiguity in the solution~\cite{sabres13}.
 Together, the transform regularizer terms $\frac{\lambda}{2} \left \| \Wb \right \|_F^2 - \lambda \log ( \, \det \Wb  \, )$ control the condition number (which is upper bounded by a monotone function of the regularizer terms~\cite{sabres13}) of the transform, with $\lambda>0$.
 Such constraints were demonstrated helpful in applications~\cite{sabres13,ravishankar2015efficient,wen2015octobos}.
 }

\subsubsection{UT-MRI}
\begin{figure}[!t] 
\begin{center} 
\vspace{-0.3in} 
\begin{tabular}{c}
\hspace{-0.1in}\includegraphics[width=5in]{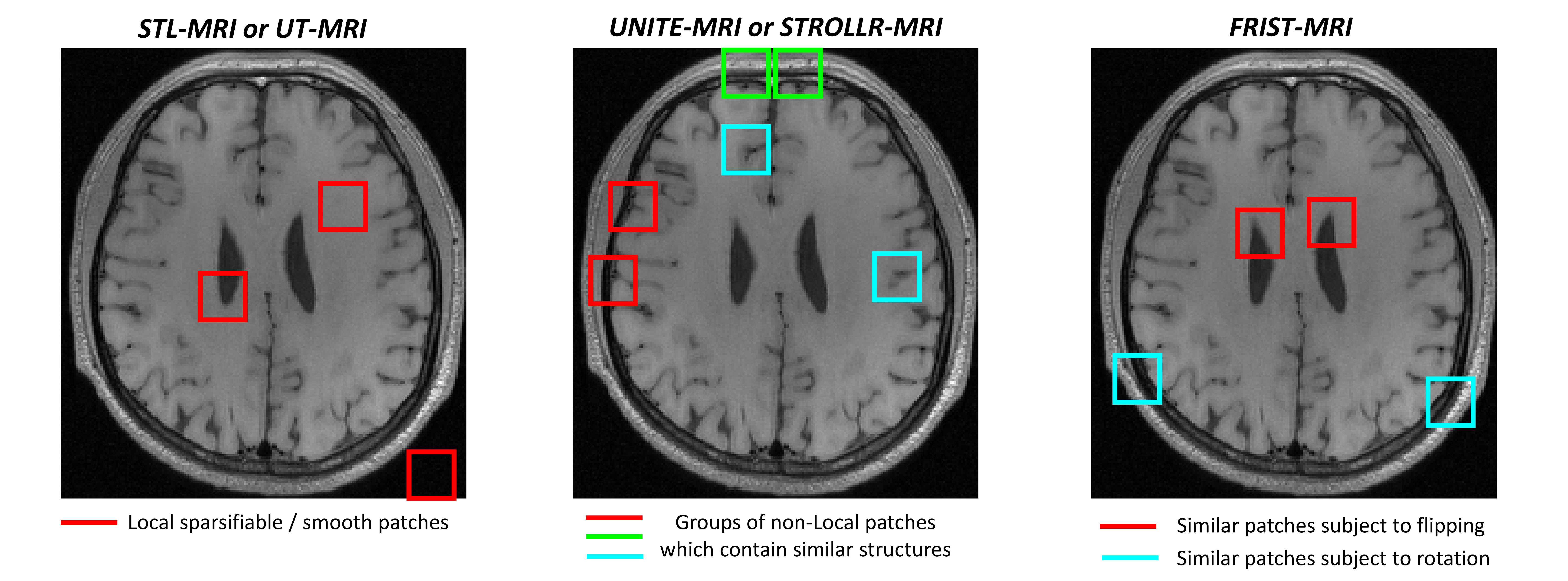} \\
\end{tabular} 
\vspace{-0.1in} 
\caption{Examples of the properties and grouping of local patches that are exploited by the variations of TL-MRI schemes.} \label{fig:property} 
\end{center} 
\vspace{-0.4in} 
\end{figure}

Instead of learning \add{generally-conditioned} transforms, 
\add{one can more efficiently} constrain the transform to be unitary (UT)~\cite{sravTCI1,wen2017frist,wen2018power}. 
This is 
akin to
$\lambda \rightarrow \infty$ in (\ref{eq:STLMRI}) of the STL-MRI scheme.
We call this variation UT-MRI, which applies the  regularizer $\Rf_{TL} (\mbf{x}) = \Rf_{UT} (\mbf{x})$ defined as
\begin{align} 
\Rf_{UT} (\mbf{x}) \triangleq \; \underset{\mbf{W}, \{ \mbf{b}_i \}}{\operatorname{argmin}} \sumi \{ \left \| \Wb \mathbf{P}_i \xb - \mbf{b}_i \right \|_2^2 + \tau^2 \left \| \mbf{b}_i \right \|_0 \} \;\;\;\;\; \text{s.t.} \;\; \; \Wb^H \Wb = \mbf{I}_n .
\label{eq:UTLMRI}
\end{align}
\add{While conventional unitary analytical transforms such as the DCT or Haar Wavelets are not data-adaptive, UT-MRI helps capture data-specific features that provide lower sparsities, thus improving performance in reconstruction.}
Both STL-MRI and UT-MRI learn a single $\Wb$ to model all image patches and mainly exploit local sparsity in the image.
We provide an example in Fig~\ref{fig:property}, showing some image patches with structures that are 
sparsified by these two schemes. \add{Other TL methods (\cite{wen2017frist} surveys some) enforce properties such as incoherence, double sparsity, etc., but have not yet been applied to MRI.}

\subsubsection{UNITE-MRI~\cite{sravTCI1}}
When MR images contain diverse features and edge information, 
it is more effective to learn multiple transforms
to model groups of ``similar'' patches.
Thus, recent work proposed 
 UNIon of Transforms lEarning (UNITE) MRI~\cite{sravTCI1}.
The regularizer $\Rf_{TL} (\mbf{x}) = \Rf_{UNITE} (\mbf{x}) $ is defined as
\begin{align} 
\nonumber \Rf_{UNITE} (\mbf{x}) \triangleq & \; \underset{\{ \mbf{b}_i \},  \{ \Wb_k, \, C_k \} }{\operatorname{argmin}} \sumk \sum_{i \in C_k} \{ \left \| \Wb_k \, \mathbf{P}_i \, \xb - \mbf{b}_i \right \|_2^2 + \tau^2 \left \| \, \mbf{b}_i \, \right \|_0 \} \\
& \;\;\;\;\;\;\;\;\;\;\;\;\;\;\;\;\;\; \text{s.t.} \;\;\; \Wb_k^H \Wb_k = \mbf{I}_n\;,\;\; \{ C_k \} \in G \;\;\;\; \forall k.
\label{eq:UNITE-MRI}
\end{align}
Here, $\{  C_k \}$ denotes a clustering of the image patches into $K$ disjoint sets, with each $C_k$ containing the indices $i$ corresponding to the patches $\{ \Wb_k \, P_i \, \xb \}_{i \in C_k}$ in the $k$-th cluster.
The superset $G$ in (\ref{eq:UNITE-MRI}) contains all possible partitions of $[1 : N ]$ into $K$ disjoint subsets.
Similar to (\ref{eq:UTLMRI}), a unitary constraint is imposed for each $\Wb_k$, which leads to an efficient 
reconstruction algorithm. When $K=1$, (\ref{eq:UNITE-MRI}) is equivalent to (\ref{eq:UTLMRI}).

The goal of UNITE-MRI is to jointly learn a union of transforms, cluster the patches based on their modeling errors with respect to each $\Wb_k$, and perform sparse coding and MR image reconstruction.
The patches ending up in the same cluster typically contain similar types of sparsifiable structures.
An example of such clustering is shown in Fig~\ref{fig:property}. In short, MR images with diverse structures are better modeled by the richer UNITE-MRI.

\subsubsection{FRIST-MRI~\cite{wen2017frist}}
MR images often
contain features that are directional (e.g., edges) and similar up to rotation and flipping.
The recently proposed FRIST-MRI scheme~\cite{wen2017frist} exploited such property of MR images using the learned Flipping and Rotation Invariant Sparsifying Transform (FRIST).
FRIST-MRI learned a \emph{parent} transform $\Wb$ such that each of its rotated and flipped \emph{children} transforms $\Wb_k = \Wb \mathbf{\Phi}_k \in \Cmb^{n \times n}$ can sparsify image patches with corresponding features. 
Here, $\{ \mathbf{\Phi}_k \}$ are the directional flipping and rotation (FR) operators~\cite{wen2017frist} that apply to each atom of $\Wb$ and approximate FR by permutation operations.
The corresponding FRIST-MRI regularizer $\Rf_{TL} (\mbf{x}) = \Rf_{FRIST} (\mbf{x})$ is defined as
\begin{align} 
\nonumber \Rf_{FRIST} (\mbf{x}) \triangleq & \; \underset{\mbf{W}, \{ \mbf{b}_i \},  \{ C_k \} }{\operatorname{argmin}} \sumk \sum_{i \in C_k} \{ \left \| \Wb \mathbf{\Phi}_k \, \mathbf{P}_i \, \xb - \mbf{b}_i \right \|_2^2 + \tau^2 \left \| \, \mbf{b}_i \, \right \|_0 \} \\
& \;\;\;\;\;\;\;\;\;\;\;\;\;\;\;\;\;\; \text{s.t.} \;\;\; \add{\Wb^H \Wb = \mbf{I}_n\;,}\;\; \{ C_k \} \in G \;\;\;\; \forall k,
\label{eq:FRIST-MRI}
\end{align}
where the clusters correspond to patches that were grouped together with a specific rotation and flip operator $\mathbf{\Phi}_k$.
Problem~\eqref{eq:FRIST-MRI} is a more structured formulation compared to \eqref{eq:UNITE-MRI} by setting $\Wb_k = \Wb \mathbf{\Phi}_k$, with only the parent transform $\Wb$ being learnable. However, a rich set of rotation and flip operators can be incorporated to generate a flexible FRIST-MRI model.
Similar to UNITE-MRI, each patch is clustered into a particular $\Wb_k$, and associated with its FR operator $\mathbf{\Phi}_k$.
The patches within the same cluster typically have similar directional features~\cite{wen2017frist}, and thus they are easier to be modeled by sparse representation.
An example of how patches can be clustered is shown in Fig~\ref{fig:property}.
Empirically, MR images with directional structures are better modeled by FRIST-MRI~\cite{wen2017frist}.

\subsubsection{STROLLR-MRI~\cite{wen2018power}}
The aforementioned variations of TL-MRI schemes~\cite{sravTCI1,wen2017frist} only model the sparsity of MR image patches, which is a strictly local image property.
However, MR images can also have non-local structures, such as self-similarity between regions, which are complementary to the local properties~\cite{wen2018power}.
Recent work proposed jointly applying transform learning and low-rank approximations over non-local patch blocks for MR image reconstruction, dubbed STROLLR-MRI~\cite{wen2018power}.
The corresponding STROLLR-MRI regularizer is a weighted sum of two components as follows:
\begin{equation} \label{eq:STROLLR}
    \Rf_{TL} (\mbf{x}) = \Rf_{STROLLR} (\mbf{x}) \; \triangleq \; \gamma^{LR} \,\Rf_{LR} (\mbf{x}) \, + \, \gamma^S \, \Rf_S (\mbf{x})\; .
\end{equation}
In \eqref{eq:STROLLR}, $\Rf_{LR} (\mbf{x})$ imposes low-rankness on groups of similar patches using a matrix rank penalty as follows:
\begin{equation} 
\Rf_{LR} (\mbf{x}) = \underset{\{ \Db_{i} \}}{\operatorname{min}} \sumi \left \{ \left \| \mathbf{V}_{i} \, \mbf{x} - \Db_{i}  \right \|_{F}^{2} + \theta^2 \, \mathrm{rank}(\Db_{i})  \right \} \, ,
\label{eq:lrReg}
\end{equation}
\add{where $\mathbf{V}_{i} : \mbf{x} \mapsto \mathbf{V}_i \mbf{x} \in \Cmb^{n \times M} $ denotes a block matching (BM) operator that groups the $M-1$ patches most similar to the reference patch $\mathbf{P}_i \mbf{x}$, within a limited search window centered at $\mathbf{P}_i \mbf{x}$, and forms a matrix, whose columns are the reference patch and its matched partners (ordered by degree of match).}\footnote{\add{When the block matching based STROLLR transform is learned from a database of images, the block matching happens within each image, and thus block matching across images can be done in parallel.}} 
Each matrix $\mathbf{V}_i \xb$ is then approximated by a low-rank matrix $\Db_{i}$, by directly penalizing its matrix rank~\cite{wen2018power}, with parameter $\theta>0$.
Besides low-rankness, the other part of the regularizer \eqref{eq:STROLLR} is a sparsity penalty of the following form:
\begin{equation} \label{eq:sparReg3D}
\Rf_S (\mbf{x}) = \underset{\{ \tilde{\mbf{b}}_i \}, \Wb}{\operatorname{min}} \sumi \left \{ \left \| \Wb \, \mathbf{C}_i \mbf{x} - \tilde{\mbf{b}}_i \right \|_{2}^{2} + \tau^2 \left \| \tilde{\mbf{b}} \right \|_{0} \right \} \;\; \text{s.t.} \;\; \Wb^H \Wb = \mbf{I}_{nl} \, .
\end{equation}
Here, the operator $\mathbf{C}_i : \mbf{x} \mapsto \mathbf{C}_i \mbf{x} \in \Cmb^{nl}$ extracts the first 
$l$ columns of $\mathbf{V}_i \mbf{x}$ corresponding to $\mathbf{P}_i \mbf{x}$ and the $l-1$ patches most similar to it, and vectorizes the sub-matrix (in column lexicographical order).
Therefore, \eqref{eq:sparReg3D} learns the transform over 3D patches for 2D MRI\footnote{In general, $\mathbf{C}_i \mbf{x}$ has an extra order of dimension over $\mathbf{P}_i \mbf{x}$.}, and thus captures non-local sparsity properties in the MR image.

The goal of STROLLR-MRI is to represent the image by joint low-rank and sparse modeling over groups of similar patches.
Similar to UNITE-MRI, highly correlated patches are modeled, as illustrated in Fig~\ref{fig:property}.
But instead of joint clustering and sparse coding, BM is applied here to explicitly group non-local but similar patches, which are reconstructed by complementary sparse and low-rank models.

\vspace{-0.1in}
\subsection{TL-MRI Algorithm Pipeline}

\begin{figure}[!t] 
\begin{center} 
\begin{tabular}{c}
\hspace{-0.1in}\includegraphics[width=6in]{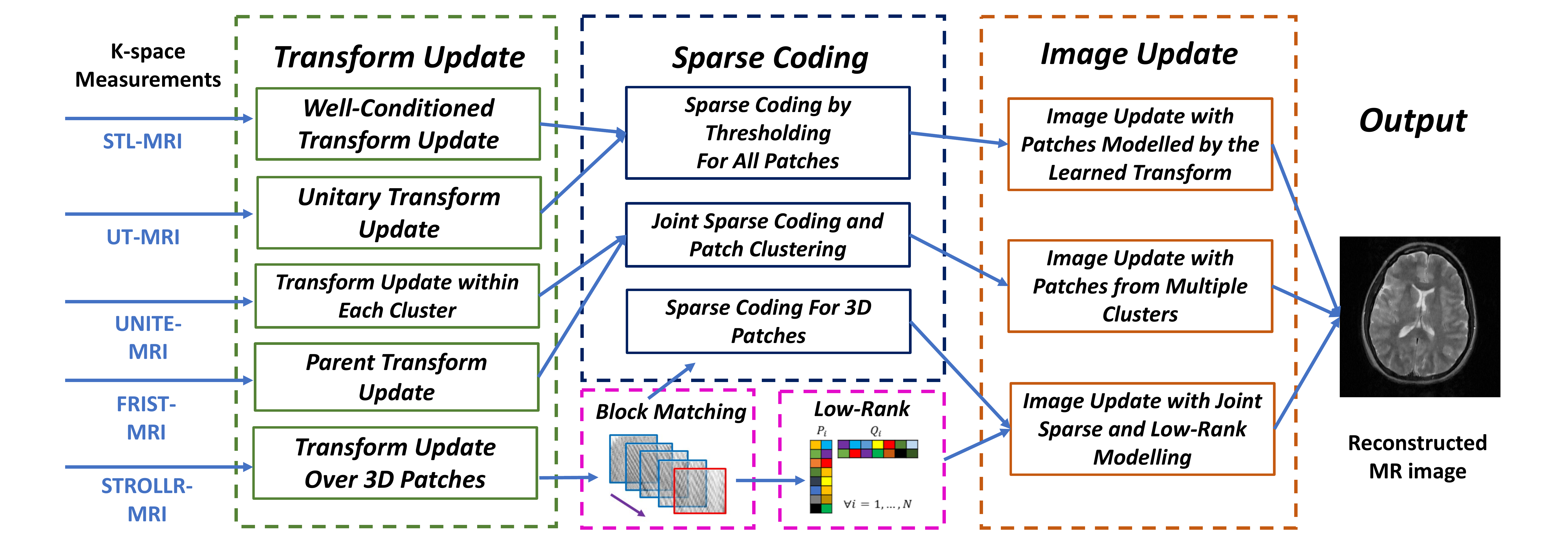} \\
\end{tabular} 
\caption{\add{A general pipeline of various TL-MRI algorithms using under-sampled k-space measurements, under the unified framework. The algorithms iterate through the transform update, sparse coding, and image update steps until convergence.}} \label{fig:pipeline} 
\end{center} 
\vspace{-0.43in} 
\end{figure}
By plugging the learnable regularizers (\ref{eq:STLMRI}) - (\ref{eq:STROLLR}) into $(\mathrm{P2})$, one can derive the MR image reconstruction algorithms of STL-MRI, UT-MRI, UNITE-MRI, FRIST-MRI, and STROLLR-MRI, respectively, that iteratively optimize the corresponding problems.
\add{These algorithms work with general acquisitions and sensing matrices, e.g., P-MRI or simple single-coil setups, }
and they are all block coordinate descent (BCD) algorithms, and involve efficient, often closed-form solutions for 
subproblems~\cite{ravishankar2015efficient,sravTCI1,wen2017frist,wen2018power}.
Though the actual forms of the regularizers (\ref{eq:STLMRI}) - (\ref{eq:STROLLR}) are varied, leading to different TL-MRI algorithms, these algorithms all contain three major steps:  (1) transform ($\Wb$) update (where transforms can often be updated in closed-form using singular value decompositions of small matrices \cite{ravishankar2015efficient,sravTCI1}); (2) generalized sparse coding; and (3) (least squares) image ($\xb$) update.
Often the sparsity penalty parameter is decreased over the algorithm iterations (a continuation strategy) for faster artifact removal initially and reduced bias over iterations \cite{sravTCI1}. 
Fig.~\ref{fig:pipeline} summarizes the aforementioned TL-MRI algorithms using a general pipeline that includes the major algorithm steps.
\add{Changes in the acquisition setup (e.g., P-MRI or simple single-coil setups) only modify the image update step in Fig.~\ref{fig:pipeline}. For example, the image update step is typically solved using FFTs for single coil Cartesian MRI, or with iterative methods such as conjugate gradients (CG) or proximal gradients for SENSE-based P-MRI.}
STROLLR-MRI involves an additional low-rank approximation step, as it jointly imposes two complementary models on the image patches~\cite{wen2018power}.
The patch grouping in STROLLR-MRI is also different from UNITE-MRI or FRIST-MRI, where it is done jointly with sparse coding. Instead, a block matching scheme is applied in STROLLR-MRI to explicitly group correlated patches~\cite{sravTCI1,wen2017frist,wen2018power}.


\vspace{-0.1in}
\subsection{Some Empirical Comparisons and Transform Visualization}

\begin{figure}[!t] 
\begin{center} 
\begin{tabular}{c}
\includegraphics[width=6in]{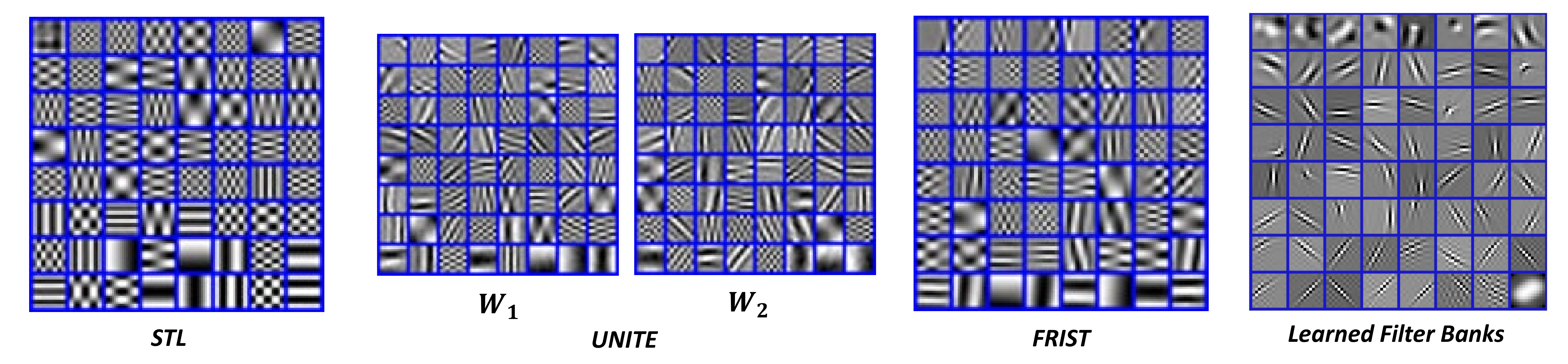} \\
\end{tabular}
\vspace{-0.1in} 
\caption{Examples of the learned transform models in the STL, UNITE (with union of two transforms), FRIST, and Filter Bank learning schemes. The atoms of each learned transform, or the impulse response of each channel in learned filter banks, are displayed as $8 \times 8$ patches.} \label{fig:visual} 
\end{center} 
\vspace{-0.4in} 
\end{figure}

\add{First,} we visualize various TL models learned on an 
MRI brain image\footnote{\add{The magnitude image was used for learning to enable simple transform visualization, but in general all TL-MRI schemes learn complex-valued transforms.}\vspace{-0.05in}}
used in ~\cite{ravishankar2015efficient}.
Fig.~\ref{fig:visual} shows the learned transform models in the STL, union of transforms, FRIST, and Filter Bank learning schemes. 
Compared to the STL model, the union of transforms model (corresponding to UNITE-MRI) in Fig.~\ref{fig:visual} captures a richer set of features in the MR image.
\add{The FRIST model has child transforms corresponding to various flipped and rotated versions of the displayed parent atoms. The generated child transforms provide a rich and flexible sparse representation model for MRI and can be particularly useful for images with many rotational features.}
Later, Section~\ref{section4a} also discusses the filterbank model shown in Fig.~\ref{fig:visual}, whose filters look quite different from the other models.


\begin{figure}[!t] 
\begin{center} 
\vspace{-0.1in}
\begin{tabular}{cccccc}
\hspace{-0.1in}
\includegraphics[width=1.in]{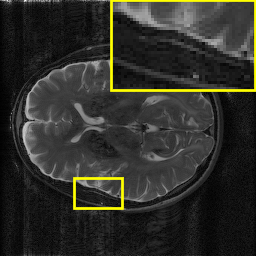} &
\includegraphics[width=1.in]{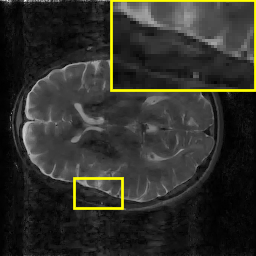} &
\includegraphics[width=1.in]{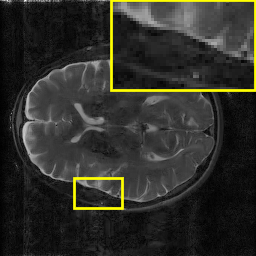} &
\includegraphics[width=1.in]{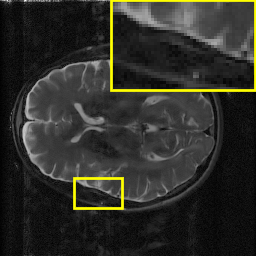} &
\includegraphics[width=1.in]{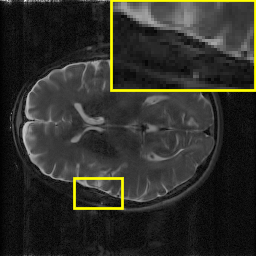} &
\includegraphics[width=1.in]{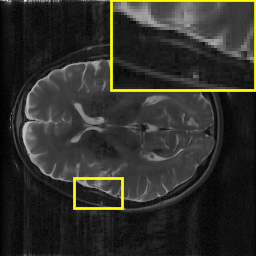}
\\
{\footnotesize Ground Truth} & 
{\footnotesize Sparse MRI} &
{\footnotesize PANO } & 
{\footnotesize DL-MRI } &
{\footnotesize STL-MRI } &
{\footnotesize STROLLR-MRI }
\\
{\footnotesize Example \textit{A}} & 
{\footnotesize ($39.07$ dB)} & 
{\footnotesize ($41.61$ dB)} & 
{\footnotesize ($41.73$ dB)} & 
{\footnotesize ($41.95$ dB) } &
{\footnotesize ($43.27$ dB) }
\\

\includegraphics[width=1.in]{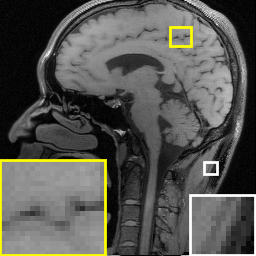} &
\includegraphics[width=1.in]{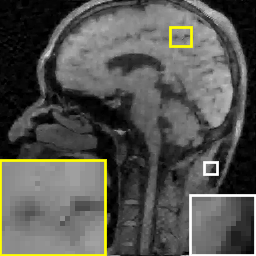} &
\includegraphics[width=1.in]{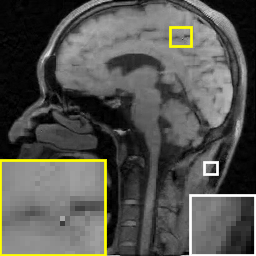} &
\includegraphics[width=1.in]{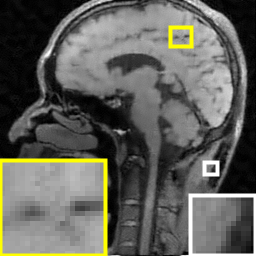} &
\includegraphics[width=1.in]{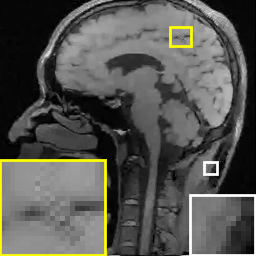} &
\includegraphics[width=1.in]{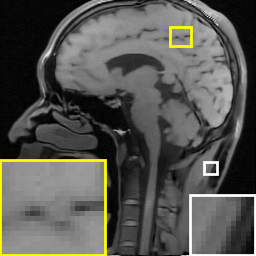}
\\
{\footnotesize Ground Truth} & 
{\footnotesize Sparse MRI} &
{\footnotesize PANO } & 
{\footnotesize DL-MRI } &
{\footnotesize ADMM-Net } &
{\footnotesize STROLLR-MRI }
\\
{\footnotesize Example \textit{B}} & 
{\footnotesize ($28.03$ dB)} & 
{\footnotesize ($30.03$ dB)} & 
{\footnotesize ($29.74$ dB)} &   
{\footnotesize ($30.67$ dB) } & 
{\footnotesize ($32.46$ dB) }
\end{tabular} 
\caption{\chg{Examples \textit{A} and \textit{B} of MRI reconstructions with simulated Cartesian 2.5x and pseudo-radial 5x undersampling, respectively, using the Sparse MRI, PANO, DL-MRI, STL-MRI, ADMM-Net, and STROLLR-MRI schemes. The reconstruction PSNRs (computed with respect to image magnitudes) are shown along with image zoom-ins. The ground truth images \textit{A} and \textit{B} are from~\cite{ravrajfes17} and \cite{sun2016deep}, respectively}.} \label{fig:mriExample}   
\end{center} 
\vspace{-0.5in} 
\end{figure}

\chg{For a simple qualitative comparison of different classes of MRI methods,
we obtain reconstructions of example MR images \textit{A} and \textit{B} shown in Fig.~\ref{fig:mriExample} from simulated single-coil $2.5$-fold undersampled (using Cartesian sampling) and 5-fold  undersampled (using pseudo-radial sampling) data, respectively.}\footnote{\chg{In Fig.~\ref{fig:mriExample}, the example \textit{A}~\cite{ravrajfes17} is a complex-valued MRI reference SENSE reconstruction (the magnitude is displayed) of $32$ channel fully-sampled Cartesian axial data from a standard spin-echo sequence; and
the example \textit{B} is a real-valued magnitude MR image, which is one of the ADMM-Net testing images provided by~\cite{sun2016deep}. For fair comparison to ADMM-Net, 
we used the same single-coil sampling mask, trained model, and testing MR image provided by the public ADMM-Net package~\cite{sun2016deep}.}}
\chg{In example \textit{A}, we reconstructed using the Sparse MRI, PANO, DL-MRI, STL-MRI, 
and STROLLR-MRI techniques, which are representative of conventional CS MRI (wavelets + total variation), partially adaptive MRI, dictionary learning, simple and advanced transform learning based methods, respectively.
In Example \textit{B}, we replaced STL-MRI by ADMM-Net~\cite{sun2016deep}, which is a representative of the deep learning methods for MRI.}\footnote{\chg{We used the parameter settings in the official packages for Sparse MRI, PANO, STL-MRI, and STROLLR-MRI. The patch size used in STL-MRI was $8\times8$ for a fair comparison to STROLLR-MRI.
The DL-MRI settings are as used in~\cite{ravrajfes17}.
DL-MRI, STL-MRI, and STROLLR-MRI follow the BCS settings.}}
Fig.~\ref{fig:mriExample} shows the reconstructions with PSNR values in decibels (dB) and local image zoom-ins.
\chg{In these  examples, the learned transforms can reconstruct higher-quality (higher PSNR) MR images than either classical Sparse MRI, or the more sophisticated PANO and DL-MRI.} 
\chg{Moreover, compared to results with transform learning (i.e., STROLLR-MRI), deep learning approaches can reconstruct images with finer details, but may also lead to more artifacts. Some similar results have been obtained in more detailed quantitative comparisons~\cite{wen2018power,sravTCI1,ravishankar2015efficient,wen2017frist}. Further comparisons with more recent additions to the rapidly evolving deep learning-based methods could be the subject of future research.}

\vspace{-0.05in}
\section{Connections to Neural Networks, New Directions, and Open Problems}
\label{section4}

This section discusses the 
connections between learned sparsifying transforms and filterbank models and reviews recent filterbank training approaches and multi-layer extensions for inverse problems. Recent progress in supervised learning of algorithms with focus on transform-based reconstruction algorithms is also discussed along with other recent trends. Finally, we discuss some of the main challenges, open problems, and future directions in the field.

\vspace{-0.1in}
\subsection{Connection to Learning Filter Banks}\label{section4a}

\add{The action of sparsifying transforms on image patches can be cast as a convolution.
Computing the inner product of a transform atom with all image patches of an image is equivalent to convolving the image with a reshaped and flipped version of the atom.
The result when picking a regularly spaced subset of image patches can be viewed as downsampling the convolved result.
With this interpretation, the sparse coefficient maps corresponding to the simple square transform model (STL) in \eqref{eq:STLMRI} are obtained by applying a set of filters to the image and thresholding the results.  This thresholding operation can be viewed as a 
certain non-linearity; namely, the proximal map of the sparsity penalty. 
Applying $\mathbf{W}^{H}$ to the sparse codes of patches and spatially aggregating them in the image (used for denoising and in the image update step of TL-MRI reconstruction algorithms) corresponds to applying matched filters to the corresponding thresholded maps and summing the results \cite{pfisbres19,ravacfess18}.
With this interpretation, the combination of patch extraction and patch-based transform operations can be viewed as applying a single transform that acts on the image as a whole.  This brings a new interpretation of overcompleteness and redundancy, as even a square (patch-based) transform becomes overcomplete when overlapping patches are used~\cite{pfisbres19}.
Recent work explored this interpretation of the transform model 
and proposed regularization schemes to learn a sparsifying filter bank that is well-conditioned and invertible at the image level, but not necessarily the patch level~\cite{pfisbres19}.}


%


\vspace{-0.2in}
\subsection{Physics-driven Deep Training of Transform-based Reconstruction Algorithms}
Recent work has considered supervised learning methods, 
wherein the underlying reconstruction model (such as a deep convolutional neural network for denoising) is learned to minimize the error in reconstructing a corpus of training images from undersampled measurements. 
\cite{leeye18,schlemper18}
These methods can require large training datasets to optimize billions of algorithm parameters (e.g., filters).
Separate works \cite{ravcfess17,ravacfess18} have cast existing 
transform learning-based blind CS image reconstruction algorithms as 
physics-driven deep convolutional networks learned on-the-fly from measurements. These networks often involve far fewer training parameters.

\begin{figure}[!t]
\hspace{-0.0in}\includegraphics[width=1.0\textwidth]{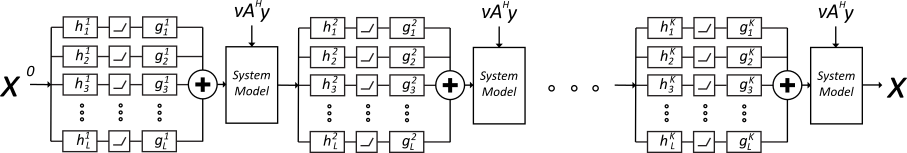}
\caption{The reconstruction model (with $K$ layers) based on the image update step of UT-MRI \cite{sravTCI1,ravacfess18}, with a general measurement operator $\mathbf{A}$ (here, $\mathbf{A}=\mbf{F}_u$). Each layer involves a decorruption step that uses filtering and thresholding operations (corresponding to transform model with $L$ filters) followed by a system model block that performs a least squares-type image update enforcing the imaging measurement model.}
\label{fig:physicsdrivenlayers}
\vspace{-0.3in}
\end{figure}

For example, the solution of the unconstrained least squares image update step of the UT-MRI algorithm \cite{sravTCI1} (which learns a unitary transform) derived from its normal equation can be represented as one layer of the model shown in Fig.~\ref{fig:physicsdrivenlayers}.
In particular, the image update step involves a system block that solves the normal equation by matrix inversion or iterative conjugate gradients (CG). This system block incorporates the forward model (physics) of the imaging process. Its inputs correspond to the right hand side of the normal equation, with $\nu \mathbf{F}_{u}^{H}\mathbf{y}$ denoting a fixed \emph{bias} term and the other input is a decorrupted version of the image after passing it through a convolutional network. First, the image is passed through a set of transform filters, followed by (hard or soft) thresholding, and then a set of matched synthesis filters, whose outputs are summed to produce a decorrupted image. The $K$ layer model in Fig.~\ref{fig:physicsdrivenlayers} corresponds to repeating the image update step for $K$ iterations in UT-MRI, with fresh filters in each iteration.
Since UT-MRI does not require training data, it can be interpreted as learning the filters of the network on-the-fly from measurements.
\add{Recent works~\cite{ravcfess17,ravacfess18} learned the filters in this multi-layer model (a block coordinate descent or BCD-Net \cite{cfess18}) using a greedy 
scheme to minimize the error in reconstructing a training set from CS measurements.} 
\add{This approach and similar approaches~\cite{sun2016deep,hammernik2018learning}} involving unrolling of MRI inversion algorithms are better referred to as \emph{physics-driven deep training} due to the systematic incorporation of the measurement model in the convolutional architecture. 
Once learned, the reconstruction architecture can be applied to test data using efficient convolution and thresholding operations and least squares-type updates. 
\add{While~\cite{ravcfess17,ravacfess18} did not enforce the corresponding synthesis and transform filters 
to be matched  
in each layer, recent work \cite{cfess18} explored learning matched filters, which may further improve image quality.}


\vspace{-0.2in}
\subsection{New Directions: Multi-Layer Transform Learning and Online Learning}

\add{Several} new directions in transform and dictionary learning for image reconstruction have been proposed recently.
For example, 
\add{various} works explored multi-layer extensions and interpretations of synthesis dictionary~\cite{multisulamelad} and sparsifying transform models~\cite{saibrenmulti}.

\begin{figure}[!t] 
\begin{center} 
\begin{tabular}{c}
\hspace{-0.1in}\includegraphics[height=1.638in]{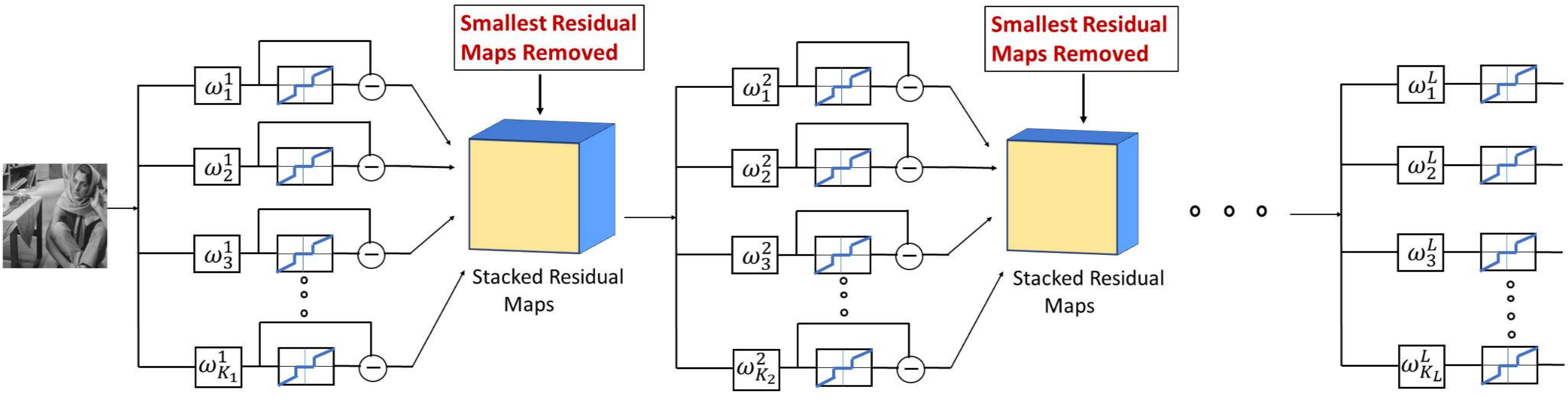} \\
\end{tabular} 
\caption{A schematic of \add{a multi-layer Transform Model \cite{saibrenmulti} (encoder) showing the filtering, thresholding, residual (difference between filter outputs and their sparse versions) generation, and downsampling (pooling) operations} in each layer.} \label{fig:multilayerTL} 
\end{center} 
\vspace{-0.4in} 
\end{figure}

\add{An efficient algorithm for unsupervised (model-based) learning of a multi-layer Deep Residual Transform (DeepResT) model (see Fig.~\ref{fig:multilayerTL}) was proposed in~\cite{saibrenmulti}.
While in conventional transform learning, we minimize the residual between the filter (or feature) maps (transformed or filtered image or patches) and their sparse (thresholded) versions (e.g., the first term in~\eqref{eq:STLMRI}), in the DeepResT model in Fig.~\ref{fig:multilayerTL}, the residual maps for different filters are stacked to form a residual volume and further jointly sparsified in the next layer,
with the goal of better minimizing the residuals over layers. 
To avoid dimensionality explosion, each filtering in each layer in Fig.~\ref{fig:multilayerTL} happens only spatially 
along each residual map channel followed by summing the results across the channels.
Moreover, to achieve robustness to noise and data corruptions, the residual volumes are downsampled (pooled) prior to further filtering. In~\cite{saibrenmulti}, for the denoising application, pooling was performed in the channel direction, where the residual maps with the smallest energies (that contain mostly noise) were dropped in each layer.
The filters and sparse maps in all layers of the DeepResT (encoder) model were jointly learned from image(s) in~\cite{saibrenmulti} to minimize the sparse residual in the final layer (layer $L$ in Fig.~\ref{fig:multilayerTL}).
Unlike conventional deep learning that uses task-based costs, the DeepResT model is learned using a transform learning (or model-based) cost in an unsupervised manner, and moreover the efficient learning algorithm in~\cite{saibrenmulti} generalized the STL method~\cite{sabres13}.}

An image encoded using the \add{learned DeepResT model (Fig.~\ref{fig:multilayerTL}) was decoded} by a linear process involving backpropagating the coefficient and residual maps through the layers.
A noisy image may be denoised by learning a multi-layer transform model directly from it and applying the follow-up decoding. This approach provided notable denoising quality improvements over single-layer learned transforms and dictionaries.  Another approach called stacking, where multi-layer models are learned again on denoised images for improved denoising, and is equivalent to successively stacking multi-layer encoder-decoder pairs, led to further improvements in denosing at high noise levels.
Ongoing work is exploring the efficacy of \add{such} multi-layer \add{learned} models for medical image reconstruction.

Another research direction that is garnering 
increasing interest is online (or time-sequential) learning for image reconstruction. 
\add{Recent works showed the promise of online transform and synthesis dictionary  learning for video denoising~\cite{wensaibresvidosat19} and dynamic MRI reconstruction from limited k-t space data~\cite{briansairajjeff18}.
Such methods could be potentially used in applications such as interventional imaging, or could also be used for effective locally-adaptive offline processing of large-scale data.}
\add{In~\cite{briansairajjeff18},}
the objective function \add{for reconstruction} changes over time and is defined as a weighted average (over time) of instantaneous cost functions corresponding to individual frames or groups (mini-batches formed using sliding windows) of consecutive frames. The instantaneous cost functions include both a data-fidelity cost and regularization based on a learned model. An exponential forgetting factor (weighting) controls the past memory (or degree of dynamic adaptivity) in the objective. Only the most recent frame(s) are optimized at a given time in the time-average cost, and the model (e.g., dictionary or transform) is adapted therein based on both the past and new information in the most recent frame measurements. Each frame is typically reconstructed in multiple windows and the estimates from those windows are (weighted) averaged to generate the final frame estimates. The online algorithm updates the model and frames in a runtime and memory efficient manner by using warm starts (initializations) for variables from estimates in previous windows (times), and only storing past information in small matrices updated cumulatively over time. By allowing the image model to change over time, online learning allows flexibility in modeling and tracking dynamically changing data. 
\add{The methods in~\cite{briansairajjeff18} provided better image quality than conventional joint or batch reconstruction at a fraction of the computational cost.
}
\chg{For example, for online learning of
a unitary sparsifying model for dynamic MRI reconstruction, \cite{briansairajjeff18} reported runtimes of 2.1 seconds to reconstruct
each frame using an unoptimized Matlab implementation.}
\add{Obtaining good trade-offs between the complexity or richness of learned models, runtime (to attain potential real-time and high performance), convergence, and dynamic adaptivity for online learning algorithms is an important direction for future MRI research.}

\vspace{-0.1in}
\subsection{Open Problems and Future Directions in Learning-Based MRI}

There are several challenges and open questions in the field of learning-driven imaging.
First, while several transform learning-driven reconstruction algorithms have been proposed, usually with guaranteed convergence to critical points (or generalized stationary points) of the problems, theory for guaranteeing recovery of the underlying image, and the properties of the filters, cost functions, and initializations to enable guaranteed and stable image recovery require further investigation. Such theory would also aid the development of better models, regularizers, and learning and reconstruction algorithms.

Second, while both unsupervised or model-based (including on-the-fly learning and online methods) and supervised learning approaches for CS MRI reconstruction have been proposed, a rigorous understanding of the relative merits of both approaches, their generalizability, and the regimes (e.g., in terms of signal-to-noise ratio or degree of undersampling) where they are each effective is needed. 
For example, supervised learning methods are highly adaptive to the distribution of the training data, and since anomalies are much less likely to occur in a training corpus than normal features, the learned models are unlikely to accurately capture them. As a consequence,   the reconstructions they produce could miss anomalies, which are all-important in medical diagnostics. This is an instance of supervised methods often not generalizing well to new instances, and is also related to the known problem of class imbalance in training data. 
\chg{On the other hand, transform learning approaches optimizing model-based costs can capture quite general properties of image sets and generalize well~\cite{yeravyongfes18spu}.}
Moreover, transform models can also be effectively and efficiently learned on-the-fly from single-instance measurements, whether or not this instance includes an anomaly.  Developing methods that combine benefits of both supervised and unsupervised or model-based learning to effectively adapt to both representative training data and anomalous instances is an interesting line of research.

Third, there has been growing interest in adaptive sampling techniques for CS MRI reconstruction, where the undersampling strategy is learned from training data to provide high quality reconstructions for specific reconstruction algorithms, including learning-based schemes \cite{saibresadaptsampl13}.
Since the underlying problems for learning the undersampling strategy are highly non-convex and combinatorial, only approximate algorithms have been proposed thus far.
Developing computationally efficient algorithms for learning the undersampling with performance guarantees is a fertile area for new research.

\add{Finally, given recent trends and breakthroughs in transform/dictionary learning and machine learning generally,  we expect the next generation of MR and hybrid imaging systems to go beyond compressed sensing and incorporate learning throughout the imaging pipeline.}
In particular, \emph{smart imaging systems} would continually learn from data (e.g., big datasets at hospitals or in the cloud, or from real-time patient data using online learning) to acquire limited measurements rapidly and efficiently perform image reconstruction and analytics. The underlying algorithms could learn or update the models for all components of the imaging process (acquisition, reconstruction, analytics) to optimize end-to-end performance in clinical settings. 
The development of such systems will require
sustained innovation in 
various models and 
efficient algorithms,  
along with innovations in pulse sequence design and hardware for incorporating learning throughout the system.

\vspace{-0.15in}
\section{Conclusions} \label{section5}

This paper briefly reviewed the timeline of compressed sensing for MRI and discussed in particular some of the advances in dictionary and transform learning for MR image reconstruction.
The sparsifying transform model enables efficient and effective sparse coding, learning, and reconstruction algorithms. We discussed a general learning-based regularization framework for MRI reconstruction. In this setting, we discussed a variety of adaptive transform regularizers based on clustering, rotation invariance, and patch similarity.  Learned sparsifying transforms are closely related to filterbanks and neural networks. Recent works have focused on physics-driven deep learning of reconstruction algorithms, unsupervised learning of multi-layer transform models, and online transform/dictionary learning.  We discussed several existing challenges in this domain and ongoing directions such as a more rigorous understanding of the pros and cons of unsupervised and supervised learning approaches for reconstruction and the development of smart imaging systems.

 \bibliographystyle{./IEEEtran}
\bibliography{./tlmri}

\begin{thebibliography}{10}
\providecommand{\url}[1]{#1}
\csname url@samestyle\endcsname
\providecommand{\newblock}{\relax}
\providecommand{\bibinfo}[2]{#2}
\providecommand{\BIBentrySTDinterwordspacing}{\spaceskip=0pt\relax}
\providecommand{\BIBentryALTinterwordstretchfactor}{4}
\providecommand{\BIBentryALTinterwordspacing}{\spaceskip=\fontdimen2\font plus
\BIBentryALTinterwordstretchfactor\fontdimen3\font minus
  \fontdimen4\font\relax}
\providecommand{\BIBforeignlanguage}[2]{{%
\expandafter\ifx\csname l@#1\endcsname\relax
\typeout{** WARNING: IEEEtran.bst: No hyphenation pattern has been}%
\typeout{** loaded for the language `#1'. Using the pattern for}%
\typeout{** the default language instead.}%
\else
\language=\csname l@#1\endcsname
\fi
#2}}
\providecommand{\BIBdecl}{\relax}
\BIBdecl

\bibitem{pMRI-Survey}
K.~P. Pruessmann, ``Encoding and reconstruction in parallel {MRI},'' \emph{NMR
  in Biomedicine}, vol.~19, no.~3, pp. 288--299, 2006.

\bibitem{setsompop16}
K.~{Setsompop}, D.~A. {Feinberg}, and J.~R. {Polimeni}, ``Rapid brain {MRI}
  acquisition techniques at ultra-high fields,'' \emph{NMR in Biomedicine},
  vol.~29, no.~9, pp. 1198--1221, 2016.

\bibitem{Fen-PT97}
P.~Feng, ``Universal spectrum blind minimum rate sampling and reconstruction of
  multiband signals,'' Ph.D. dissertation, University of Illinois at
  Urbana-Champaign, Mar 1997, {Y}oram Bresler, adviser.

\bibitem{donoho2006compressed}
D.~L. Donoho, ``Compressed sensing,'' \emph{IEEE Trans. Information Theory},
  vol.~52, no.~4, pp. 1289--1306, 2006.

\bibitem{lustig2007sparse}
M.~Lustig, D.~Donoho, and J.~M. Pauly, ``Sparse {MRI}: The application of
  compressed sensing for rapid {MR} imaging,'' \emph{Magnetic Resonance in
  Medicine}, vol.~58, no.~6, pp. 1182--1195, 2007.

\bibitem{sai2011dlmri}
S.~Ravishankar and Y.~Bresler, ``{MR} image reconstruction from highly
  undersampled k-space data by dictionary learning,'' \emph{IEEE transactions
  on medical imaging}, vol.~30, no.~5, pp. 1028--1041, 2011.

\bibitem{sravTCI1}
------, ``Data-driven learning of a union of sparsifying transforms model for
  blind compressed sensing,'' \emph{IEEE Transactions on Computational
  Imaging}, vol.~2, no.~3, pp. 294--309, 2016.

\bibitem{sun2016deep}
Y.~Yang, J.~Sun, H.~Li, and Z.~Xu, ``Deep {ADMM-Net} for compressive sensing
  {MRI},'' in \emph{Advances in Neural Information Processing Systems}, 2016,
  pp. 10--18.

\bibitem{ravcfess17}
S.~{Ravishankar}, I.~Y. {Chun}, and J.~A. {Fessler}, ``Physics-driven deep
  training of dictionary-based algorithms for {MR} image reconstruction,'' in
  \emph{2017 51st Asilomar Conference on Signals, Systems, and Computers},
  2017, pp. 1859--1863.

\bibitem{hammernik2018learning}
K.~Hammernik, T.~Klatzer, E.~Kobler, M.~P. Recht, D.~K. Sodickson, T.~Pock, and
  F.~Knoll, ``Learning a variational network for reconstruction of accelerated
  {MRI} data,'' \emph{Magnetic Resonance In Medicine}, vol.~79, no.~6, pp.
  3055--3071, 2018.

\bibitem{sabres13}
S.~Ravishankar and Y.~Bresler, ``Learning sparsifying transforms,'' \emph{IEEE
  Transactions on Signal Processing}, vol.~61, no.~5, pp. 1072--1086, 2013.

\bibitem{wen2018power}
B.~Wen, Y.~Li, and Y.~Bresler, ``The power of complementary regularizers: Image
  recovery via transform learning and low-rank modeling,'' \emph{IEEE
  Transactions on Image Processing}, 2019, to appear.

\bibitem{wensaibresvidosat19}
B.~{Wen}, S.~{Ravishankar}, and Y.~{Bresler}, ``{VIDOSAT}: High-dimensional
  sparsifying transform learning for online video denoising,'' \emph{IEEE
  Transactions on Image Processing}, vol.~28, no.~4, pp. 1691--1704, 2019.

\bibitem{lingal13}
S.~G. Lingala and M.~Jacob, ``Blind compressive sensing dynamic {MRI},''
  \emph{IEEE Trans. Medical Imaging}, vol.~32, no.~6, pp. 1132--1145, 2013.

\bibitem{ye2002self}
J.~C. Ye, Y.~Bresler, and P.~Moulin, ``A self-referencing level-set method for
  image reconstruction from sparse fourier samples,'' \emph{International
  Journal of Computer Vision}, vol.~50, no.~3, pp. 253--270, 2002.

\bibitem{akccakaya2011low}
M.~Ak{\c{c}}akaya, T.~A. Basha, B.~Goddu, L.~A. Goepfert, K.~V. Kissinger,
  V.~Tarokh, W.~J. Manning, and R.~Nezafat, ``Low-dimensional-structure
  self-learning and thresholding: regularization beyond compressed sensing for
  mri reconstruction,'' \emph{Magnetic Resonance in Medicine}, vol.~66, no.~3,
  pp. 756--767, 2011.

\bibitem{liu2015balanced}
Y.~Liu, J.-F. Cai, Z.~Zhan, D.~Guo, J.~Ye, Z.~Chen, and X.~Qu, ``Balanced
  sparse model for tight frames in compressed sensing magnetic resonance
  imaging,'' \emph{PloS one}, vol.~10, no.~4, p. e0119584, 2015.

\bibitem{liu2016projected}
Y.~Liu, Z.~Zhan, J.-F. Cai, D.~Guo, Z.~Chen, and X.~Qu, ``Projected iterative
  soft-thresholding algorithm for tight frames in compressed sensing magnetic
  resonance imaging,'' \emph{IEEE transactions on medical imaging}, vol.~35,
  no.~9, pp. 2130--2140, 2016.

\bibitem{qu2012undersampled}
X.~Qu, D.~Guo, B.~Ning, Y.~Hou, Y.~Lin, S.~Cai, and Z.~Chen, ``Undersampled
  {MRI} reconstruction with patch-based directional wavelets,'' \emph{Magnetic
  Resonance Imaging}, vol.~30, no.~7, pp. 964--977, 2012.

\bibitem{qu2014magnetic}
X.~Qu, Y.~Hou, F.~Lam, D.~Guo, J.~Zhong, and Z.~Chen, ``Magnetic resonance
  image reconstruction from undersampled measurements using a patch-based
  nonlocal operator,'' \emph{Medical image analysis}, vol.~18, no.~6, pp.
  843--856, 2014.

\bibitem{liang2007spatiotemporal}
Z.-P. Liang, ``Spatiotemporal imaging with partially separable functions,'' in
  \emph{2007 4th IEEE International Symposium on Biomedical Imaging: From Nano
  to Macro}.\hskip 1em plus 0.5em minus 0.4em\relax IEEE, 2007, pp. 988--991.

\bibitem{ravmoorerajfes17}
S.~{Ravishankar}, B.~E. {Moore}, R.~R. {Nadakuditi}, and J.~A. {Fessler},
  ``Low-rank and adaptive sparse signal ({LASSI}) models for highly accelerated
  dynamic imaging,'' \emph{IEEE Transactions on Medical Imaging}, vol.~36,
  no.~5, pp. 1116--1128, 2017.

\bibitem{haldar2014low}
J.~P. Haldar, ``Low-rank modeling of local $k$-space neighborhoods ({LORAKS})
  for constrained {MRI},'' \emph{IEEE transactions on medical imaging},
  vol.~33, no.~3, pp. 668--681, 2014.

\bibitem{ravishankar2015efficient}
S.~Ravishankar and Y.~Bresler, ``Efficient blind compressed sensing using
  sparsifying transforms with convergence guarantees and application to
  magnetic resonance imaging,'' \emph{SIAM Journal on Imaging Sciences},
  vol.~8, no.~4, pp. 2519--2557, 2015.

\bibitem{pfisbres19}
L.~{Pfister} and Y.~{Bresler}, ``Learning filter bank sparsifying transforms,''
  \emph{IEEE Transactions on Signal Processing}, vol.~67, no.~2, pp. 504--519,
  2019.

\bibitem{saibrenmulti}
S.~{Ravishankar} and B.~{Wohlberg}, ``Learning multi-layer transform models,''
  in \emph{2018 56th Annual Allerton Conference on Communication, Control, and
  Computing (Allerton)}, 2018, pp. 160--165.

\bibitem{wen2017frist}
B.~Wen, S.~Ravishankar, and Y.~Bresler, ``{FRIST}- flipping and rotation
  invariant sparsifying transform learning and applications,'' \emph{Inverse
  Problems}, vol.~33, no.~7, p. 074007, 2017.

\bibitem{ravrajfes17}
S.~{Ravishankar}, R.~R. {Nadakuditi}, and J.~A. {Fessler}, ``Efficient sum of
  outer products dictionary learning ({SOUP-DIL}) and its application to
  inverse problems,'' \emph{IEEE Transactions on Computational Imaging},
  vol.~3, no.~4, pp. 694--709, 2017.

\bibitem{wang2016accelerating}
S.~Wang, Z.~Su, L.~Ying, X.~Peng, S.~Zhu, F.~Liang, D.~Feng, and D.~Liang,
  ``Accelerating magnetic resonance imaging via deep learning,'' in \emph{2016
  IEEE 13th International Symposium on Biomedical Imaging (ISBI)}.\hskip 1em
  plus 0.5em minus 0.4em\relax IEEE, 2016, pp. 514--517.

\bibitem{schlemper18}
J.~{Schlemper}, J.~{Caballero}, J.~V. {Hajnal}, A.~N. {Price}, and
  D.~{Rueckert}, ``A deep cascade of convolutional neural networks for dynamic
  {MR} image reconstruction,'' \emph{IEEE Transactions on Medical Imaging},
  vol.~37, no.~2, pp. 491--503, 2018.

\bibitem{leeye18}
D.~{Lee}, J.~{Yoo}, S.~{Tak}, and J.~C. {Ye}, ``Deep residual learning for
  accelerated {MRI} using magnitude and phase networks,'' \emph{IEEE
  Transactions on Biomedical Engineering}, vol.~65, no.~9, pp. 1985--1995,
  2018.

\bibitem{wang2018image}
G.~Wang, J.~C. Ye, K.~Mueller, and J.~A. Fessler, ``Image reconstruction is a
  new frontier of machine learning,'' \emph{IEEE Transactions on Medical
  Imaging}, vol.~37, no.~6, pp. 1289--1296, 2018.

\bibitem{cfess18}
I.~Y. {Chun} and J.~A. {Fessler}, ``Deep {BCD}-net using identical
  encoding-decoding cnn structures for iterative image recovery,'' in
  \emph{2018 IEEE 13th Image, Video, and Multidimensional Signal Processing
  Workshop (IVMSP)}, 2018, pp. 1--5.

\bibitem{zhan2016fast}
Z.~Zhan, J.-F. Cai, D.~Guo, Y.~Liu, Z.~Chen, and X.~Qu, ``Fast multiclass
  dictionaries learning with geometrical directions in {MRI} reconstruction,''
  \emph{IEEE Transactions on Biomedical Engineering}, vol.~63, no.~9, pp.
  1850--1861, 2016.

\bibitem{ravacfess18}
S.~{Ravishankar}, A.~{Lahiri}, C.~{Blocker}, and J.~A. {Fessler}, ``Deep
  dictionary-transform learning for image reconstruction,'' in \emph{2018 IEEE
  15th International Symposium on Biomedical Imaging (ISBI 2018)}, 2018, pp.
  1208--1212.

\bibitem{wen2015octobos}
B.~Wen, S.~Ravishankar, and Y.~Bresler, ``Structured overcomplete sparsifying
  transform learning with convergence guarantees and applications,''
  \emph{International Journal of Computer Vision}, vol. 114, no. 2-3, pp.
  137--167, 2015.

\bibitem{multisulamelad}
J.~{Sulam}, V.~{Papyan}, Y.~{Romano}, and M.~{Elad}, ``Multilayer convolutional
  sparse modeling: Pursuit and dictionary learning,'' \emph{IEEE Transactions
  on Signal Processing}, vol.~66, no.~15, pp. 4090--4104, 2018.

\bibitem{briansairajjeff18}
B.~E. Moore, S.~Ravishankar, R.~R. Nadakuditi, and J.~A. Fessler, ``Online
  adaptive image reconstruction ({OnAIR}) using dictionary models,'' \emph{IEEE
  Transactions on Computational Imaging}, 2019, to appear.

\bibitem{yeravyongfes18spu}
S.~Ye, S.~Ravishankar, Y.~Long, and J.~A. Fessler, ``{SPULTRA:} low-dose {CT}
  image reconstruction with joint statistical and learned image models,''
  \emph{IEEE Transactions on Medical Imaging}, 2019, to appear.

\bibitem{saibresadaptsampl13}
S.~{Ravishankar} and Y.~{Bresler}, ``Adaptive sampling design for compressed
  sensing {MRI},'' in \emph{2011 Annual International Conference of the IEEE
  Engineering in Medicine and Biology Society}, 2011, pp. 3751--3755.

\end{thebibliography}
\end{document}